\documentclass[12pt]{iopart}

\usepackage{graphicx}
\usepackage{float} 
\usepackage{natbib}
\graphicspath{ {./images/} }

\begin{document}

\title[]{GWOPS: A VO-technology driven tool to search for the electromagnetic counterpart of gravitational wave event }

\author{Yunfei Xu$^1$$^,$$^2$$^,$$^3$, Dong Xu$^1$,Chenzhou Cui$^1$$^,$$^3$,  Dongwei Fan$^1$$^,$$^3$, Zipei Zhu$^4$$^,$$^1$,  Bangyao Yu$^1$, Changhua Li$^1$$^,$$^3$, Jun Han$^1$$^,$$^3$, Linying Mi$^1$$^,$$^3$, Shanshan Li$^1$$^,$$^3$,  Boliang He$^1$$^,$$^2$$^,$$^3$, Yihan Tao$^1$$^,$$^3$,  Hanxi Yang$^1$$^,$$^3$, \& Sisi Yang$^1$$^,$$^3$,  }
\address{$^1$  National Astronomical Observatories, Chinese Academy of Sciences, Beijing 100101, PR China}
\address{$^2$ University of Chinese Academy of Sciences, Beijing 100049, PR China}
\address{$^3$ National Astronomical Data Center, Beijing 100101, PR China}
\address{$^4$ Huazhong University of Science and Technology, Wuhan 430074, China}
\ead{ccz@nao.cas.cn,dxu@nao.cas.cn}
\vspace{10pt}
\begin{indented}
\item[]July 2020
\end{indented}

\begin{abstract}
The search and follow-up observation of electromagnetic (EM) counterparts of gravitational waves (GW) is a current hot topic of GW cosmology. Due to the limitation of the accuracy of the GW observation facility at this stage, we can only get a rough sky-localization region for the GW event, and the typical area of the region is between 200 and 1500 square degrees. Since GW events occur in or near galaxies, limiting the observation target to galaxies can significantly speedup searching for EM counterparts. Therefore, how to efficiently select host galaxy candidates in such a large GW localization region, how to arrange the observation sequence, and how to efficiently identify the GW source from observational data are the problems that need to be solved. International Virtual Observatory Alliance (IVOA) has developed a series of technical standards for data retrieval, interoperability and visualization. Based on the application of VO technologies, we construct the GW follow-up Observation Planning System (GWOPS). It consists of three parts: a pipeline to select host candidates of GW and sort their priorities for follow-up observation, an identification module to find the transient from follow-up observation data, and a visualization module to display GW-related data. GWOPS can rapidly respond to GW events. With GWOPS, the operations such as follow-up observation planning, data storage, data visualization, and transient identification can be efficiently coordinated, which will promote the success searching rate for GWs EM counterparts.
\end{abstract}

%
\vspace{2pc}

\section{Introduction}

On August 17, 2017, the LIGO-Virgo Collaboration (LVC) gravitational wave (GW) detector network captured a GW signal from the spin of two compact star remnants (neutron stars). Only 1.7 seconds after the observation of this signal, a gamma-ray burst named GRB170817A \citep{connaughton2017grb170817a} was detected by the Fermi gamma-ray burst monitoring telescope. The GW and gamma-ray alerts were sent to the astronomical community to launch a series of follow-up observations. Global astronomers eventually detected the decaying electromagnetic (EM) signals from the neighboring NGC4993 galaxy. The multi-messenger observation with EM and GW facilities marks a new era in the multi-messenger and time-domain astronomy \citep{abbott2017multi}

GW sources can be roughly divided into three types: continuous GW sources, stochastic GWs, and burst GW. Continuous GW sources, such as fast-rotating neutron stars, emit quasi-sinusoidal GWs over durations much longer than the detectors' lifetime. Stochastic GWs can take the form of a cosmological background, similar to the EM cosmic microwave background, or could arise from a cacophony of GW sources at closer distances. Burst GW are transient signals that have not been well modeled or unknown waveforms. Burst sources include the supernova, the merger and post-merger phases of merging compact binaries. Compact binary coalescence (CBC) is a binary system that rotates with each other, where either or both constituents are a black hole or neutron star. CBCs are the best characterized and one of the most promising sources for the advanced detectors, with a realistic expected rate of 20 such events per year observed at design sensitivity \citep{abbott2018prospects}.

As a significant tool for astrophysics and cosmology research, multi-messenger joint observations have important applications in GW cosmology study. GW sources (such as CBC) can be used as a standard siren to measure the geometry and expansion of the universe, and from this to test cosmological models, measure cosmological parameters, and study dark energy and dark matter \citep{del2012inference}. As the EM waveband observation is the only way to determine the GW source's location and redshift, it is necessary to complete the joint detection of the EM signals produced by the GW source for finding the astrophysical origin of GW or conducting further research on their physical properties. Therefore, from the perspective of GW astronomy, the significance of the observation of EM counterparts of GWs can be compared to the direct detection of GW signals.

However, the ground GW observation network has poor positioning ability to GW source, and it is only possible to obtain a rough sky localization of the event. For the most typical binary neutron star (BNS) system, a single detector is impossible to achieve positioning. When the number of detectors is increased to two, the positioning of the source of GWs can be limited to one ring zone. When three or more detectors are connected, the localization region of the GW source can be reconstructed. In general, the elliptical localization region of the GW source is about 200\textasciitilde1500 square degrees. Even the aLIGO A+ detector network upgraded after 2020 has not improved much on positioning ability \citep{abbott2018prospects}. Due to the different sensitivity and maximum detection distance of GW detectors (aLIGO \citep{harry2010gm} 190Mpc, AdVirgo \citep{acernese2014advanced} 125Mpc, KAGRA \citep{aso88kagra} 140Mpc), even if all planned detectors are built and used in the future, the localization region will still be large for GW events that exceed the detection range of AdVirgo and KAGRA. Besides, the per-year BNS search volume increases giving an expected up to 180 BNS detections annually. The high detection rate of GWs and the low accuracy of the positioning have brought great challenges to the follow-up observation of the EM counterpart. It requires new methods to search for them.

\cite{gehrels2016galaxy} proposed a strategy to search for GW EM counterparts based on galaxy observations. The strategy assumes that sources of GWs reside within (or close to) the normal matter seen as galaxies. With the host galaxy's information, the estimation of the localization region of the GW event can be improved. Based on the strategy, we can retrieve the galaxies which locate in the GW localization range and follow-up observe them to find if any transients there. This strategy has been widely adopted and further studied. such as \citep{arcavi2017optical}, \citep{chan2017maximizing}, and \citep{ducoin2020optimizing}. After summarizing these works, we believe that building an efficient and automated software system for GW follow-up observation collaboration based on the galaxy observation strategy, is essential to improve the efficiency of GW EM counterpart's search.  In the system the issues need to be solved are how to rapidly select host candidates in such a vast GW localization region, and how to sort the observation priorities of them, as well as how to identify the GW source from observational data efficiently. In response to these issues, a series of technical difficulties need to be resolved. In order to rapidly respond to GW events, the galaxies contained in the GW localization region need to be found out from the database in real-time, which requires an appropriate data retrieval method. Besides, for the identification of transients, except the observation data itself, additional data are needed to be retrieved to assist in the identification, including the light curve data of known transients in the region,  and observation data of the target area from other survey projects, etc. Moreover, a unified organization, management and visualization framework for observational data is also needed to enable observing scientists to understand the data of GW follow-up observations more intuitively. Virtual Observatory (VO) \citep{cui_zhao_2007} specifies a series of technical standards for data retrieval, interoperability and visualization, which provide technical solutions for working out these issues. In this paper, we introduce a GW follow-up Observation Planning System, referred to as GWOPS for short\footnote{GWOPS can be accessed at: https://nadc.china-vo.org/gwops/.}, which is constructed base on VO technologies. GWOPS has a pipeline to automatically process GW alerts and filter the host candidates as follow-up targets, and it also consists of the components to realize the identification of the EM counterpart of GW and data visualization. 

In Section 2, we briefly introduce the overview of this system, including the system architecture, data stream, and the database schema. Section 3 describes the components of GWOPS in detail, especially the VO-related technologies in host galaxy filtering, transient identification and data visualization. Section 4 presents the current status and the future work of the system. GWOPS has supported the follow-up observation of GW events for six optical telescopes in China during the third observing run (O3) of LVC. It also has good scalability to provide other telescopes around the world with the support of follow-up observation of GW events. We conclude in Section 5 with a discussion.

\section{System Overview}
\subsection{System architecture and Data Processing Stream}

\begin{figure}
\centering 
\includegraphics[width=0.8\textwidth]{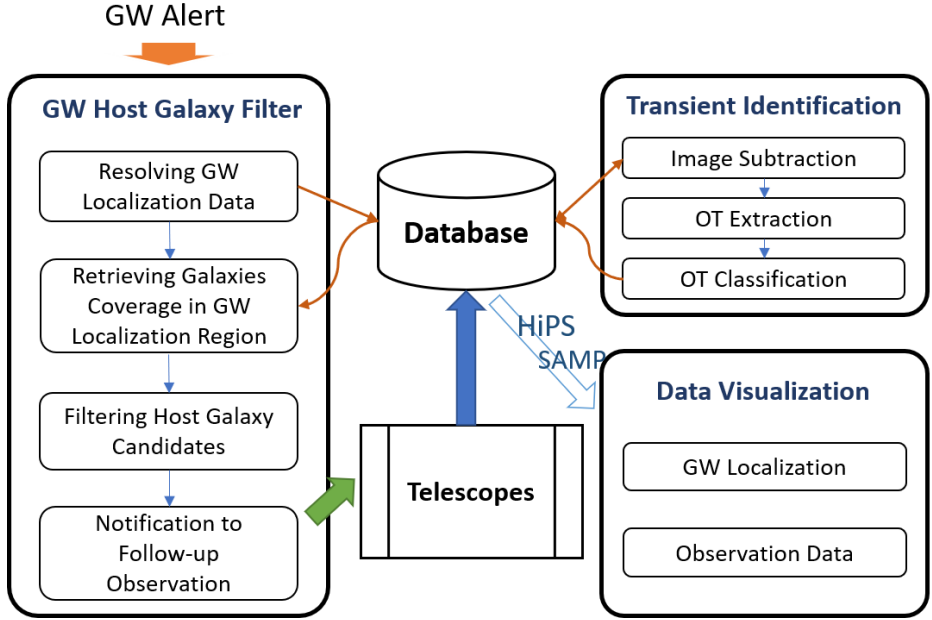} 
\caption{ The architecture and data processing stream of GWOPS. The three components of GWOPS implement collaboration through data interaction. The starting point of the data flow is the GW alert. After processing by the GW host galaxy filter, the follow-up observation target list is sent to telescopes, then data generated by follow-up observations are stored in the database. Afterward, the transient identification and visualization component read observation data and GW event data from the database for processing, and finally displaythem to users. Many functions of GWOPS are implemented in the backend, such as host galaxy filtering and extraction of transient candidates, which are transparent to users. The frontend web portal provides users with data display and interaction functions. } 
\label{Fig.1}
\end{figure}

GWOPS consists of three parts, the GW host galaxy filter, the transient identification, and the data visualization component. As Fig.\ref{Fig.1} shows, the GW host galaxy filter monitors the GW event alert released by LVC via the NASA Gamma-Ray Coordinates Network (GCN) \citep{barthelmy2016gcn}. After receiving the alert, it automatically downloads the GW localization data from the gravitational-wave candidate event database (GraceDB)\footnote{GraceDB: https://gracedb.ligo.org/}. Based on the localization data, the galaxies database is queried to obtain all the host candidates in the localization region.  The candidates are further filtered and sorted. Then the top 150 of the ranking are sent to telescopes for follow-up observation, with all the observation data storing in the database. The transient identification component is used to identify candidates of GW EM counterpart from the observation targets (OT), which are extracted from observation data by image recognition method. The processes of host galaxy filtering  and OT candidates extraction runs in the backend of GWOPS and are transparent to users. Registered users can check the original observation data by logging into the GWOPS web portal, and can manually identify the OT candidates there. Non-registered users can view the Hierarchical Progressive Survey (HiPS) \citep{fernique2017hips}  generated based on the observation data, as well as the follow-up observation list on the  GWOPS visualization component, which is web-based and provides intuitive data visualization effect. The original observation data and transient lists will be released to non-registered users before the start of the LVC O4 run.

\subsection{Database design}
\begin{figure}
\centering 
\includegraphics[width=1\textwidth]{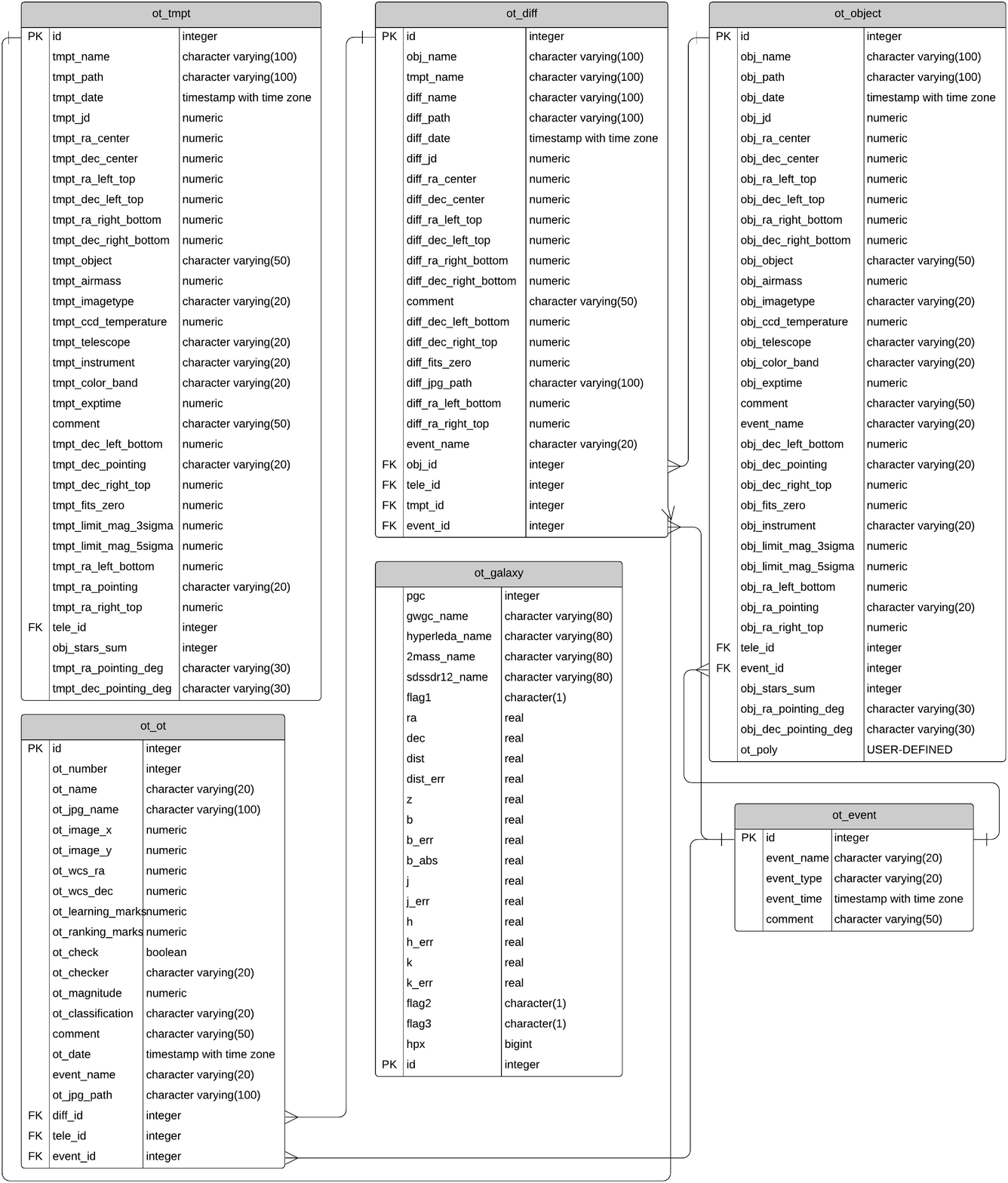} 
\caption{GWOPS database schema. The database is implemented with the open source database software PostgreSQL \footnotemark[1]. Each table uses the self-increasing ID as the primary key, and the foreign key as the association between tables. Each table has a B-tree index on ID column to improve retrieval efficiency.} 
\label{Fig.2}
\end{figure}

Based on the data obtained by follow-up observations, we have established a GW observation database. The core data tables in the database are the Event table, Galaxy table, Observation table, Template table, Diff table, and OT table. The schema of the database is shown in Fig.\ref{Fig.2}.
\begin{enumerate}
\item Event table: It records the ID, issued time, estimated distance, and localization region of each GW event. 

\item Galaxy table: It stores galaxy information in the range of 200MPC, including each galaxy’s location, distance, redshift, as well as B-band luminosity, based on the Galaxy List for the Advanced Detector Era (GLADE) catalog \citep{dalya2018glade}, which version is 2.3. GLADE is a fusion of galaxy data through multiple catalogs, specifically used for GW host galaxy search and the filtering of EM counterpart target for follow-up observation. The five galaxy catalogs used to obtain GLADE are GWGC \citep{white2011list}, 2MPZ \citep{bilicki2013two}, 2MASS XSC \citep{jarrett20002mass}, HyperLEDA \citep{prugniel2005hyperleda} and SDSS DR12Q \citep{kozlowski2017vizier}, and they complement each other in the wavelength and galaxy types. GLADE contains a total of 3.6 million galaxies and is complete up to \~{}37Mpc, has a completeness of \~{}61 percent within the maximal value of single-detector BNS ranges for aLIGO during O2 (\~{}100 Mpc), \~{}54 percent within the minimal planned BNS range during O3, and \~{}48 percent within the planned BNS range of single aLIGO detectors with design sensitivity (\~{}173 Mpc) \citep{dalya2018glade}.

\item Observation table: This table records the observation images for each GW event.

\item Template table: It contains the information of previously observed images as the templates for extracting OTs.

\item Diff table: It stores secondary data derived from the image subtraction of the observation image and the template image.

\item OT table: The table records information of OTs. 
\end{enumerate}

\footnotetext[1]{https://www.postgresql.org/}

\section{GWOPS Components}

To respond rapidly to GW events and find EM counterparts efficiently and conveniently, GWOPS is meant to be a real-time follow-up observation planning pipeline,  an efficient transient identification tool, and a user-friendly observation data browser. In this section, we will describe the components of GWOPS and the VO technology applied in them. 

\subsection{GW Host Galaxy Filter}

\begin{figure}
\centering 
\includegraphics[width=1\textwidth]{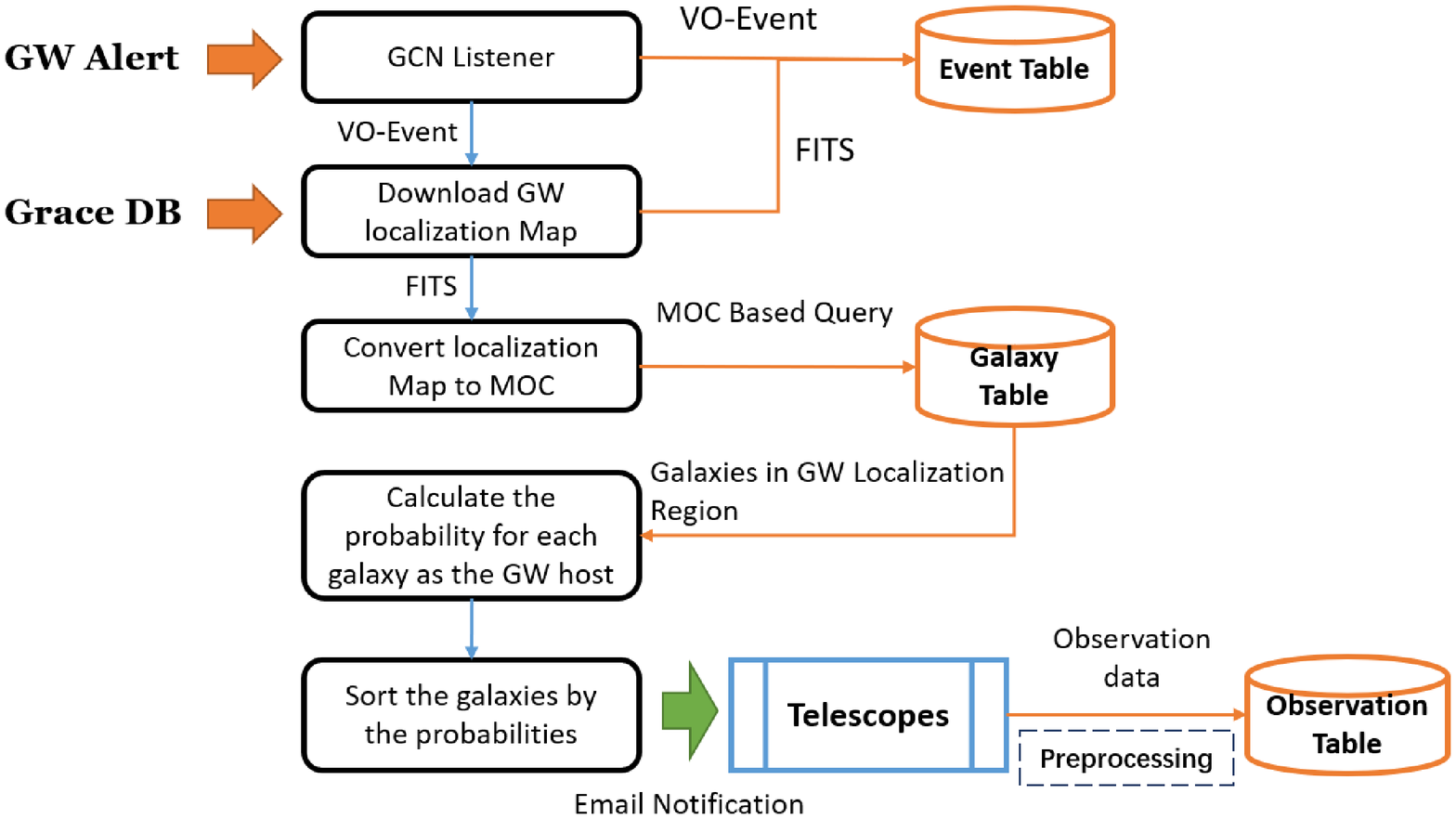} 
\caption{The dataflow of GW Host Galaxy Filter. The external data of GW Host Galaxy Filter comes from GW Alert of LVC, and the GW localization data from GraceDB. The data output of the component is the follow-up observation list, which flows out to the follow-up observation telescopes.} 
\label{Fig.3}
\end{figure}

The GW host galaxy filter component selects the potential GW host galaxies to reduce the search range in the sky. We implement filtering in two steps. The first step is to retrieve the galaxies covered in the localization region of the GW event. However, the number of galaxies we get in this step is around 10,000 \~{} 50,000. For the operability of follow-up observation, we calculate each galaxy's probability as the GW host and sort the observation list by their probabilities. These processes are completed automatically at the backend of GWOPS, and users can check the detailed information of each alert and the list of follow-up observations at the frontend, i.e., the web portal. Fig. \ref{Fig.3} shows the data flow of the components.

\subsubsection{Galaxies retrieving by GW localization region}

A VO-Event alert will be sent out by LVC through the GCN when a candidate GW event is detected. The alert contains a localization map with distance constraints. In general, the GW localization region is an irregular strip, but the general spatial search is a cone query or a polygonal query. So it is necessary to find out how to use the localization region as a query condition to retrieve galaxies from the database. The GW localization map is a FITS file, which records the event occurrence probability at each HEALPix (the Hierarchical Equal Area isoLatitude Pixelization) \citep{gorski2005healpix} grid. We convert the FITS file into a MOC (Multi-Order Coverage Map) file \citep{fernique2014moc} to speed up the query. The MOC is an International Virtual Observatory Alliance (IVOA) standard that used to specify arbitrary sky regions, and its mechanism is based on the HEALPix sky tessellation algorithm.

Then we build the spatial index for the galaxy data table in the database. For each entry in the data table, a HEALPix ID at the order 10 can be calculated by the coordinates of the entry. Then a clustered index is built on the HEALPix ID column, which can make the entries that are nearby in space are also nearby on disk. The MOC file represents a range of sky regions by a collection of HEALPix indices at different orders. The recursive nature of the HEALPix indexing scheme leads to the property that all grids at a higher order have indices consecutive between two values prescribed by the lower order. Therefore, we convert all the indices which order lower than 10 into the two values at order 10. Then a request for galaxy entries in a MOC leads to a query for galaxies whose indices are between these two values. The nature of the ordering on the disk means that all of these galaxy entries can be found in consecutive blocks or pages on disk, which leads to fast access, requiring only a few disk reads.

\subsubsection{Sorting candidates for follow-up observation}

It is impossible to fully observe so many galaxies in the localization region within a limited time, so it's necessary to filter the host candidates and prioritize their observations. In GWOPS, a Bayesian inference based method \citep{fan2014bayesian} is adopted to calculate the probability that a galaxy is the GW event host, then sorted the galaxies by the probability. The method uses the galaxy's physical characteristics (B-band brightness, distance) and the probability of the GW event occurred at the galaxy’s location as the prior, the posterior probability of each galaxy to be the host of the GW event is calculated. Eqn. \ref{eq1} is applied to calculate the posterior probability.
\begin{eqnarray}
p(\gamma |D,S,M)=\frac{p(\gamma|S,M)p(D|S,\gamma,M)}{p(D|S,M)} \label{eq1}
\end{eqnarray}

 The meaning of the parameters:
\begin{itemize}
\item $\gamma$: the common parameters observed by the GW event and the host galaxy, specifically referring to the position parameters of the galaxy in the localization region, $\alpha$ (RA), $\beta$ (Dec), d (Distance)

\item D: observation data of GW event 

\item S: observation data of EM

\item M: host galaxy candidate information
\item p(D $\vert$ S, $\gamma$, M) is the likelihood probability of the data set D given S, $\gamma$, i.e., the probability of the corresponding position of galaxies given in the GW localization map
\item  p($\gamma$ $\vert$ S, M) is the prior probability formula
\end{itemize}

The prior probability formula is as Eqn. \ref{eq2}:
\begin{eqnarray}
\fl p(\gamma |S, M)\propto (\frac{D_{gc}}{D_{GW}})^{3}\frac{1}{N} \sum_{j=1}^{N}\delta (\alpha -\alpha_{j}, \beta-\beta_{j}, d-d_{j})L_{B_{j}}+\frac{3L_{Bmean}}{4\pi D^3_{GW}}H(d-D_{gc})d^2  \label{eq2}
\end{eqnarray}

Where $N$ denotes the total number of galaxies in the galaxy data table. $D_{gc}$ denotes the range of the galaxy data table. $D_{GW}$ denotes the detection range of the GW detector. $L_{B_{j}}$ denotes the luminosity of the host galaxy in $B$ band, and $H$ denotes a step function.

\cite{fan2014bayesian} have verified this method based on the simulation data of dense binary-star merger. The results show that in the range of 200Mpc, in the 8.5\% simulation results, the probability that the top 10 galaxies in the ranking are GW event host galaxies is 50\%. In 10\% of the simulations, the host galaxy is indeed located in the top 10 galaxies.

\subsubsection{Transient Identification}

\begin{figure}
\centering 
\includegraphics[width=0.9\textwidth]{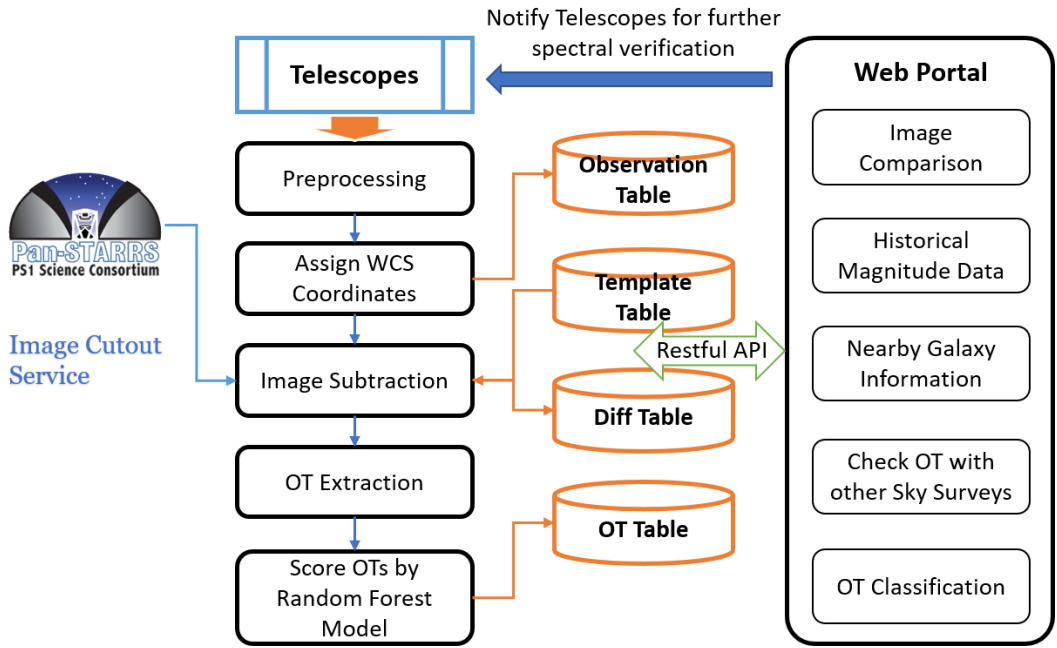} 
\caption{ The data flow and modules of the transient identification component. The backend of the component interacts with the frontend (web portal) through data retrieves and Restful APIs. The frontend displays the processing results of the backend to the user, and assists the user to identify the transient with the auxiliary data.} 
\label{Fig.4}
\end{figure}

The transient identification component extracts OTs based on the residuals between the observation data and the template data in the backend, and provides users with an operation interface for manually identifying OTs in the frontend. The data flow and modules of the component are shown in Fig. \ref{Fig.4}.

The workflow in transient identification components is as follows:

\begin{enumerate}
\item Preprocessing: The observation data returned by the telescope are subjected to a series of processing, including reducing the flat field and background to make it a clean image. 
\item Assign World Coordinate System (WCS) coordinates \citep{mink1997wcstools}: In this process, the WCS coordinate information is added into the observation data based on the coordinate of observation targets. This is the basis for observation data retrieval and visualization. After completing this process, the observation data is stored in the observation data table.
\item Image subtraction: The OTs are obtained by subtracting the observation data from the template data. Template data mainly comes from two sources. One is the historical image taken by the observation telescope and stored in the template table. The second is to use the image cut out service \citep{haridas2005fits} in the IVOA's Simple Image Access (SIA) protocol \citep{dowler2015ivoa} to obtain data taken by other surveys. The image cut out service we used extracts rectangular regions of some larger image, returning an image of the requested size to the client. Such images are usually drawn from a database or a collection of survey images that cover some large portion of the sky. GWOPS gets the template data from Pan-STARRS \citep{kaiser2002pan} in the absence of historical template data.
\item OT extraction: The SExtractor \citep{bertin1996sextractor} is used to extract OTs from observation data, template data and residual data automatically. Then the position and brightness of the OTs obtained from the three are compared. OTs with significant brightness changes will be recorded and transferred to the next processing stage.
\item Score OTs for manual identification: Base on the  investigation of the astronomical sources identification methods in PTF/iPTF/ZTF \citep{mahabal2019machine}, Pan-STARRS1 \citep{wright2015machine} and Subaru/Hyper Suprime-Cam survey \citep{lin2018machine}, we proposed a trained random forest model to evaluate the extracted OTs and give a score of the OT as a real transient. Only OTs with a score higher than 0.2 (maximum 1.0) will be listed for manual identification, and this score will also be used as a reference for manual identification.
\end{enumerate}

\begin{figure}
\centering 
\includegraphics[width=0.9\textwidth]{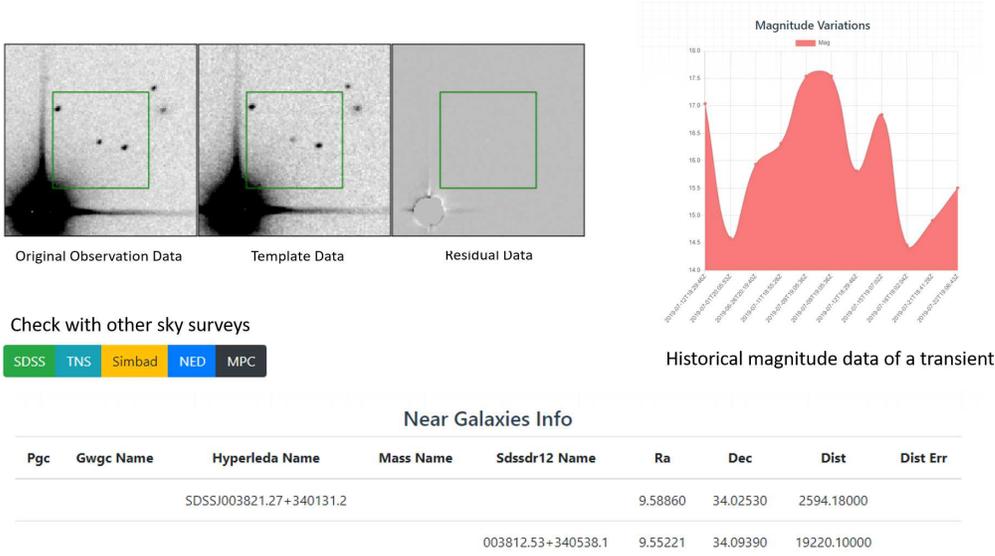} 
\caption{ Transient identification support information displayed in the web portal. The upper left picture shows an OT original observation image, template image and residual image in turn. By clicking the buttons below the picture, users can view the data of the corresponding survey or database at the OT location. The line chart on the right is the historical magnitude data of the OT. The table below is the information of the galaxies near the OT, and the galaxy data comes from GLADE V2.3} 
\label{Fig.5}
\end{figure}

After the automatic processing above, the high-scoring OTs will be listed on the web portal for manual verification. Besides, OTs related information will also be listed on the webpage for reference (as shown in Fig. \ref{Fig.5}), including the observation data, template data, residual data of each OT, and its historical magnitude data, as well as the information on nearby galaxies of the OT. Besides, we also use the IVOA simple cone search protocol  \citep{williams2011ivoa} to check whether other sky surveys find transient near the position of the OT.  From the web portal, we can check if there is any sources in the OT’s position in SDSS, TNS (Transient Name Server) \footnote{TNS: https://wis-tns.weizmann.ac.il/}, Simbad \citep{wenger2000simbad}, NED (NASA/IPAC Extragalactic Database)\citep{helou1991nasa} and MPC (Minor Planet Center) \citep{marsden1980minor}. Finally, based on this information, we can determine whether the OT is a real transient and classify its type. The real transient will be sent to telescopes for further spectral verification.

\subsection{Visualization}

\begin{figure}
\centering 
\includegraphics[width=0.9\textwidth]{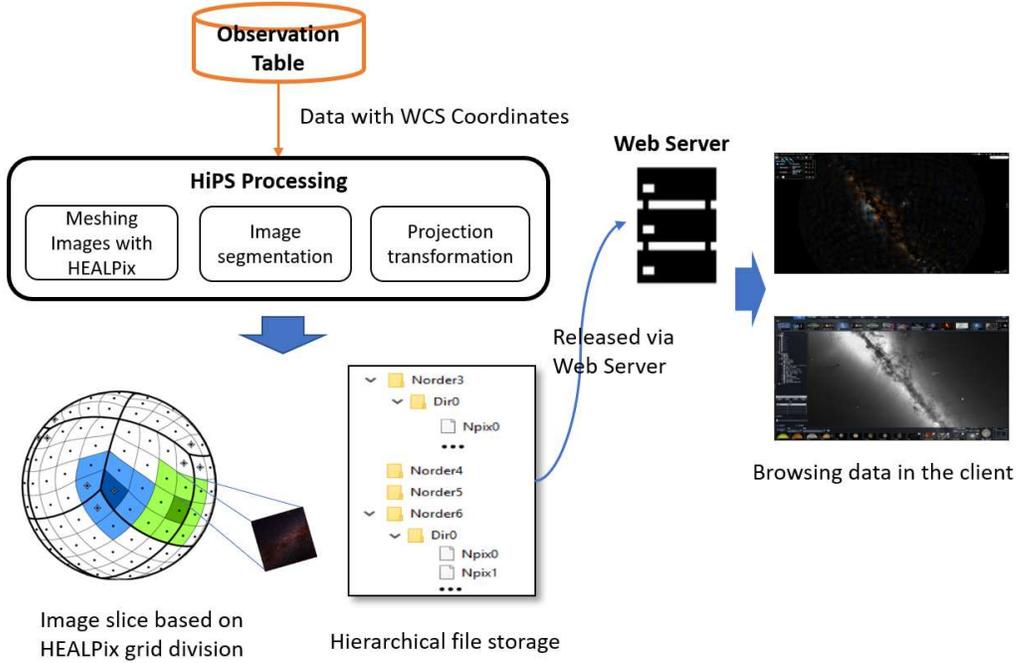} 
\caption{ The original observation data is transformed into HiPS data through HEALPix grid mapping, image segmentation, and projection conversion. The HiPS data are stored in the file system in PNG format and indexed by the HEALPix number. The data are then published through a web server, and clients that support HiPS (such as Aladin and WWT) can load the HiPS link to browse the data.} 
\label{Fig.6}
\end{figure}

The visualization component allows users to view the observation data in the GW localization region intuitively. GWOPS uses Aladin Lite \citep{boch2014aladin} as the facility for data visualization on the web.  Aladin Lite is a WebGL-based Javascript library. It can display FITS images loaded WCS coordinates on the three-dimensional celestial sphere, and draw contours on the celestial sphere based on JSON files, which are resolved by GW localization data. The observation data is converted into IVOA HiPS for visualization. As shown in Fig. \ref{Fig.6}, HiPS uses the HEALPix spherical segmentation method to mesh the observation data, convert the data from the original projection to the HEALPix projection, and recursively divides the data into image pyramids of different resolutions according to the hierarchical division of HEALPix. 

\begin{figure}
\centering 
\includegraphics[width=0.8\textwidth]{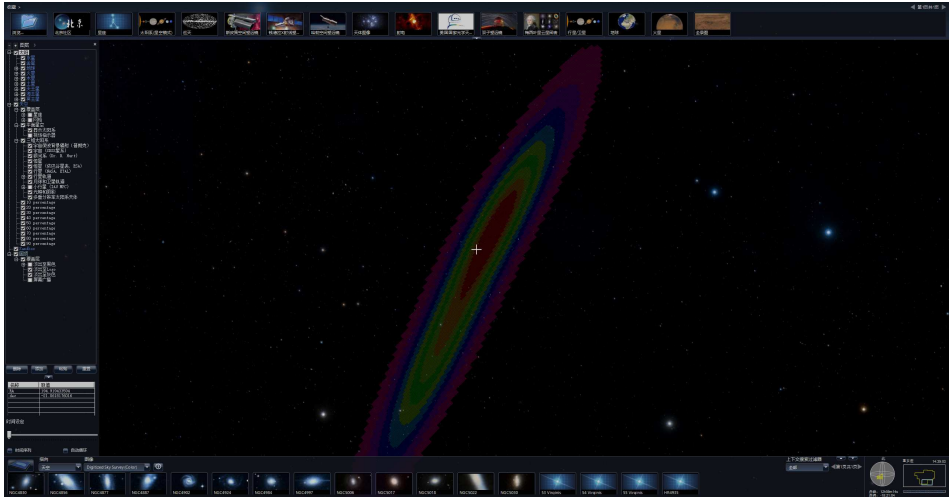} 
\caption{ The visualization of GW localization data (GW170817) in WWT. Different colors represent different confidence regions.} 
\label{Fig.7}
\end{figure}

The data in GWOPS can also be visualized in WorldWide Telescope (WWT) \citep{rosenfield2018aas}.  This requires the user to additionally install the China-VO version of Worldwide Telescope \footnote{China-VO WWT can be downloaded at http://wwt.china-vo.org} on the computer and start it in advance. WWT is a scientific data visualization platform launched in 2008 by Microsoft Research and then open-sourced in 2015. We have successfully implemented the HiPS schema in WWT \citep{xu2020ivoa}, which make the observation data can be browsed freely on it. As for the visualization of GW localization data, the sky region file is converted into a series of key-value pairs files with HEALPix grid and positioning probability. These files are encapsulated as VOTable, and users can transmit the files to WWT via the IVOA SAMP protocol \citep{taylor2015samp} by clicking the 'Send to WWT' button on the data browser page. The visualization of GW localization data in WWT is shown in Fig. \ref{Fig.7}.

We chose WWT as one of the visualization facilities not only because it has excellent visualization effect and supports a series of VO data access standards, but also because it has the unique feature of Guided Tour, which is a kind of recorded path with narration, text, and imagery. With this feature, users can create tours with the GW data to share their discovery with other researchers or the public, and further promote GW-related research.

\section{Project Status and Future Work}

Currently, six telescopes are using GWOPS to run their follow-up observation for the GWs EM counterpart searching. The telescopes are located in Nanshan Xinjiang, Xuyi Jiangsu, and Xinglong Hebei of China. Information about these telescopes is listed in Table \ref{tab1}.
\def\degree{${}^{\circ}$}
\begin{table}[]
\caption{\label{tab1}Telescopes supported by GWOPS during LVC O3 run}
\footnotesize\rm
\begin{tabular*}{\textwidth}{@{}l*{15}{@{\extracolsep{0pt plus 12pt}}l}}
\br
\textbf{Telescope} & \textbf{Location}    & \textbf{Altitude} & \textbf{FOV} & \textbf{CCD Size} & \textbf{Aperture/m} \\
\mr
\textbf{HMT}       & 88.5772E    45.5224N & 1795              & 60’ x 40’    & 4008 x 2672       & 0.5                 \\
\textbf{CNEOST}    & 11.46E        32.73N & 210.3             & 3\degree x 3\degree      & 10560 x 10560     & 1.2/1               \\
\textbf{Schmidt}   & 117.9640E  40.6508N  & 950               & 1.5\degree x 1.58\degree & 4096 x 4096       & 0.6/0.9             \\
\textbf{NOWT}      & 87.1777E    43.4708N & 2080              & 1.3\degree x 1.3\degree  & 4160 x 4136       & 1                   \\
\textbf{NEXT}      & 87.1777E    43.4708N & 2080              & 20’ x 20’    & 2048 x 2064       & 0.6                 \\
\textbf{XL216}     & 117.9640E  40.6508N  & 950               & 10” x 10”    & 2048 x 2048       & 2.16             \\
\br  
\end{tabular*}
\end{table}

The LVC has officially started the third observing run on April 1, 2019 \citep{ligoo3run}. As of March 2020, 53 GW events have been found \citep{farr2020latest}. GWOPS responded to each event promptly. We recorded the response efficiency of several events, which are shown in table \ref{tab2}. Thanks to the construction of an efficient spatial index, GWOPS is able to complete the selection of host candidates in the GW localization region within 500ms and can send follow-up observation lists to telescopes less than 30 seconds after receiving an alert.

\begin{table}[]
\caption{\label{tab2}The response efficiency of GWOPS to GW events,  where LDT denotes the time consumption of downloading the localization data from GraceDB. LRA represents the area of the localization region. LCT is the time consumption of converting the localization data to MOC. RN refers to the number of galaxies in the localization region. HCT denotes the time consumption of retrieving the host candidates from the galaxy data table. SCT denotes the time consumption of probability calculation and sorting of the host candidates.}
\footnotesize\rm
\begin{tabular*}{\textwidth}{@{}l*{15}{@{\extracolsep{0pt plus 12pt}}l}}
\br
\textbf{GW Event} & \textbf{LDT (s)} & \textbf{LRA ($deg^{2}$)} & \textbf{LCT(s)} & \textbf{RN }  & \textbf{HCT (ms)} & \textbf{SCT(s)}\\
\mr
\textbf{S190421ar}       & 1.49 & 1444             & 17.76   & 36783       & 455.82 & 2.98                 \\
\textbf{S190426c}    & 1.52  & 1131             & 17.44  &15564 &276.31 & 1.46     \\
\textbf{S190512at}   & 1.48  & 252               & 19.16  &10967      & 166.22 &0.91             \\
\textbf{S190630ag}      & 1.48  &1486          & 17.53  &84205       & 715.84  &5.20                 \\
\textbf{S190701ah}      & 1.51   & 49            & 19.05    & 2867       &85.86 &2.95                 \\
\textbf{S190816i}     & 1.52  & 1467              &17.49     & 74432      & 659.85 &  4.78          \\
\textbf{S190910d}     & 1.46  & 2482              &16.86     &67380      &746.97 &  4.94          \\
\textbf{S191105e}     & 1.49  & 643              &18.91     &39365      &342.44 &  2.46         \\
\textbf{S191129u}     & 1.50  & 852              &17.13     &41720      &385.31 &2.75       \\
\textbf{S191204r}     & 1.49  &103              &18.36     &2938     &87.26 &2.69      \\
\textbf{S191222n}     & 1.51  &1850            &17.27     &69673     &690.73 &4.89    \\
\br  
\end{tabular*}
\end{table}

Most of the GW events found during O3 are binary black hole merger events, and only eight events are considered to be BNS events, but too far away to capture clear EM signals, so we have not been able to find any real EM counterparts of GW events.

\begin{figure}
\centering 
\includegraphics[width=0.8\textwidth]{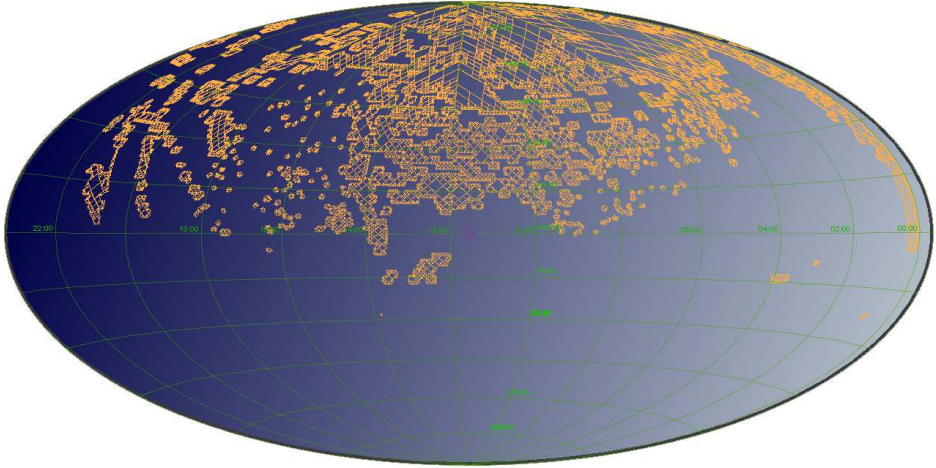} 
\caption{The coverage of observed sky regions. The yellow grid represents the observed area. Since GWOPS did not provide support for telescopes in the southern hemisphere during O3, only observations in the northern sky area and covered a total of 6,133 square degrees} 
\label{Fig.8}
\end{figure}

The telescopes, which followed these GW events, are also surveying the northern sky. As of March 2020, a total of 21,172 images have been taken, covering an area of 6,133 square degrees, as shown in Fig. \ref{Fig.8}. Based on the GWOPS transient identification component, these observations were processed in real-time, and the extracted transient candidates were manually identified.

In the next stage, we will work to turn GWOPS into an open collaboration platform. We notice that there is already a fantastic GW follow-up observation collaboration platform named Gravitational Wave Treasure Map \citep{wyatt2020gravitational}. GWOPS has common features and differences with the Treasure Map. Both of them realize the visualization of information about each GW event. The difference is that GWOPS establishes a pipeline that implements the follow-up observation planning for observation facilities, and generates observation data based on these facilities to search for electromagnetic counterparts. Treasure Map has a great openness to form a collaborative platform, and GWOPS is currently mainly serving the observation facilities that joined in GWOPS and through the open data to collaborate with other researchers. We will make a series of improvements to GWOPS to make it more open and robust, which are listed as follows:

\begin{enumerate}

\item As an independent system from the observation facility, GWOPS can also support other telescopes to run GW follow-up observation. We have developed a subscription function which can send follow-up observation lists to subscribers via email, and we are planning to implement an interface to make other telescopes  take advantage of the features provided by GWOPS, turning GWOPS into an open collaboration platform. The visualization system will also be upgraded to access multi-wavelength data from other observation facilities. Moreover, each candidate's visibility will be calculated based on the geographic location and the maximum pitch angle of each telescope, to exclude invisible sources from the observation list for a specific telescope.

\item Improve the OT automatic extraction and scoring algorithm for the transient identification component. At the current stage, the SExtractor is used to extract the OT from the image. A random forest model is used to determine whether the OT is a real transient, and a score is given. The accuracy of current algorithms is not high enough, resulting in more workload of manual verification. The pattern recognition algorithm based on deep learning is preferred to make many improvements.

\item The data retrieval method provided by VO allows us to obtain multi-wavelength data. These data can be used as an essential reference for transient identification. We will establish a multi-wavelength reference database for GWOPS transient identification. The database will cover observational data from ultraviolet to radio wavelength. While the OT identification and classification is performed manually, the known sources of other wavelengths at the target position can be obtained for auxiliary identification.

\end{enumerate}

\section{Summary}

GWOPS is a system to improve the efficiency of follow-up observation and EM counterpart identification for the GW event. The system is composed of three components: GW host galaxy filter, transient identification, and the visualization component. Based on the galaxy observations strategy, the GW host filter selects galaxies from the GW localization area as the target of subsequent observation. As an automated pipeline, the GW host filter can respond to GW alerts in real-time. With the Bayesian inference algorithm, the probability of a candidate host galaxy can be calculated. Then the top 150 galaxies sorted by probability are sent as potential targets to telescopes by email for follow-up observations. The transient identification component will automatically process the observation data and use the image recognition software to extract the possible transients. A trained random forest model is used to score the transient candidates, and the candidates with higher scores can be manually identified in a user-friendly and efficient web portal. In the visualization component, we use Aladin Lite and WWT as the facilities for users to browse the observation data and GW event related data. Users can also make Guided Tour on WWT with these data to promote their latest research and perform GW related education and public outreach.

GWOPS is a VO technology-driven system. In each component of GWOPS, there are applications of VO related technologies. In the GW host galaxy filter, the MOC standard of IVOA is used to implement the database query based on the GW localization region. In the transient identification component, the IVOA SIA standard is used to retrieve images of Pan-STARRS as template data, and the IVOA SCS standard is applied to retrieve archived data of other surveys at the location of the transient candidate. In the visualization component, the HiPS schema of IVOA is implemented for the organization and visualization of observation data, and the SAMP standard of IVOA is adopted to transfer commands and data between components.

During the O3 of LVC, GWOPS has supported six optical telescopes in China to run the GWs follow-up observation. GWOPS can respond to GW event alerts in real-time, complete the filtering of GW host candidates and submit them to the telescope within 25s. As of March 2020, a total of 6,133 square degrees of sky area was observed, and 21,172 targets were obtained. However, only 8 of the GW events during O3 are possible BNS events, and they are too far or not in the telescopes’ visible range. There are no real EM counterparts of GW events found yet by our system.

More features will be added to GWOPS to make it a robust and open collaboration platform for GW EM counterpart searching. We will provide an interface for other telescopes to utilize the features and data of GWOPS. We also try to apply deep learning methods to improve the accuracy of transient source identification. Meanwhile, a multi-wavelength reference database base on VO standards to assist in identifying the GWs EM counterpart will be integrated.

\section*{Acknowledgments}

This work is supported by National Natural Science Foundation of China (NSFC)(11803055), the Joint Research Fund in Astronomy (U1731125, U1731243, U1931132) under cooperative agreement between the NSFC and Chinese Academy of Sciences (CAS). Data resources are supported by National Astronomical Data Center (NADC) and Chinese Virtual Observatory (China-VO). This work is supported by Astronomical Big Data Joint Research Center, co-founded by National Astronomical Observatories, Chinese Academy of Sciences, and Alibaba Cloud. Dong Xu acknowledges the supports by the One-Hundred-Talent Program of the Chinese Academy of Sciences (CAS) and by the Strategic Priority Research Program “Multi-wavelength Gravitational Wave Universe” of the CAS (No. XDB23000000).

\bibliography{reference}
\end{document}